\documentclass{article}
\usepackage{spconf,amsmath,graphicx}
\usepackage{booktabs}
\usepackage{etoolbox}
\patchcmd{\thebibliography}
  {\settowidth}
  {\setlength{\itemsep}{0pt plus 0.1pt}\settowidth}
  {}{}
\apptocmd{\thebibliography}
  {\small}
  {}{}

\def\y{{\mathbf y}}

\usepackage{tikz}
\usepackage{textcomp}
\usepackage{hyperref}
\usepackage{lipsum}

\title{SINGLE-CHANNEL SPEECH ENHANCEMENT WITH DEEP COMPLEX U-NETWORKS AND PROBABILISTIC LATENT SPACE MODELS}

\name{Eike J. Nustede, Jörn Anemüller\thanks{Copyright \textcopyright 2023 IEEE. Personal use of this material is permitted.  Permission from IEEE must be obtained for all other uses, in any current or future media, including reprinting/republishing this material for advertising or promotional purposes, creating new collective works, for resale or redistribution to servers or lists, or reuse of any copyrighted component of this work in other works. Funded by Deutsche Forschungsgemeinschaft (DFG, German Research Foundation) under project ID 352015383, SFB 1330/B3.}}

\address{Carl von Ossietzky University Oldenburg, Computational Audition Group,\\
Dept. med. Physics \& Acoustics and Cluster of Excellence Hearing4all,\\
Oldenburg, Germany\\
\{eike.jannik.nustede, joern.anemueller\}@uol.de}

\begin{document}
%
\maketitle

\begin{abstract}
In this paper, we propose to extend the deep, complex U-Network architecture for speech enhancement by incorporating a probabilistic (i.e., variational) latent space model. The proposed model is evaluated against several ablated versions of itself in order to study the effects of the variational latent space model, complex-value processing, and self-attention.
Evaluation on the MS-DNS 2020 and Voicebank+Demand datasets yields consistently high performance. E.g., the proposed model achieves an SI-SDR of up to 20.2~dB, about 0.5~to~1.4~dB higher than its ablated version without probabilistic latent space, 2--2.4~dB higher than WaveUNet, and 6.7~dB above PHASEN.
Compared to real-valued magnitude spectrogram processing with a variational U-Net, the complex U-Net achieves an improvement of up to 4.5~dB SI-SDR.
Complex spectrum encoding as magnitude and phase yields best performance in anechoic conditions whereas real and imaginary part representation results in better generalization to (novel) reverberation conditions, possibly due to the underlying physics of sound.

\end{abstract}


\section{INTRODUCTION}
\label{sec:intro}

Speech enhancement algorithms are a necessity for tasks involving distorted audio signals at low signal-to-noise-ratio (SNR), such as automatic speech recognition \cite{Li2014}, voice over IP applications \cite{Harte2015TCDVoIPAR}, and speaker verification \cite{shon19b_interspeech}. Current algorithms focus on denoising by increasing signal quality via improved SNR, consequently increasing speech intelligibility for listeners.\newline
State-of-the-art approaches leverage various architectures of denoising deep neural networks, differentiated by their general architecture and the type of input used, mainly split into processing either time-domain or spectral features.
Recurrent neural networks (RNNs) focus on real-time application \cite{Xia2020}, while fully convolutional networks often yield better SNRs \cite{Luo2019} with frame-based processing. Generative models \cite{SEGAN} and, more recently, attention-based networks \cite{koizumi2020} were adapted for speech enhancement. The U-Network architecture \cite{Ronneberger2015} has successfully been adapted by several authors for speech enhancement. Dilated convolutions \cite{Bosca2020} and attention models \cite{AttentionUnet2020} have been shown to be beneficial for denoising distorted speech signals. A complex-valued U-Net was proposed \cite{choi2018phaseaware}, causing attention to shift to phase-aware networks \cite{Phasen}. Showing state-of-the-art performance, complex-valued approaches were further developed \cite{hu20g_interspeech}, including transformer-based U-Networks \cite{Uformer}. The U-Network structure is similar to an autoencoder, yet the use of probabilistic latent space models similar to variational autoencoders, was previously only used for image segmentation tasks \cite{kohl2019}.\newline
In our previous work \cite{DVunet}, we showed that including a probabilistic latent space model in a U-Network increases its generalization ability by introducing noise to the latent space, thereby indirectly increasing the processed feature variety of the network during training. Here, we extend our previous work to model the latent space of a complex-valued U-Network by introducing two separate latent spaces for real and imaginary part, respectively. The network is evaluated and compared against several ablated versions, as well as WaveUNet \cite{Stoller2018} and PHASEN \cite{Phasen}. Additionally, we compare combinations of log-scaled power and phase spectra with an input containing real and imaginary parts of the underlying complex signal, including a suitable loss function. A weighted loss function with emphasis on the phase/imaginary part is proposed to train both model variations.

\section{METHODS}
\label{sec:methods}

\subsection{Complex U-Network model}
\label{ssec:cumodel}
\begin{figure}[t]
    \centering
    \includegraphics[width=1\linewidth]{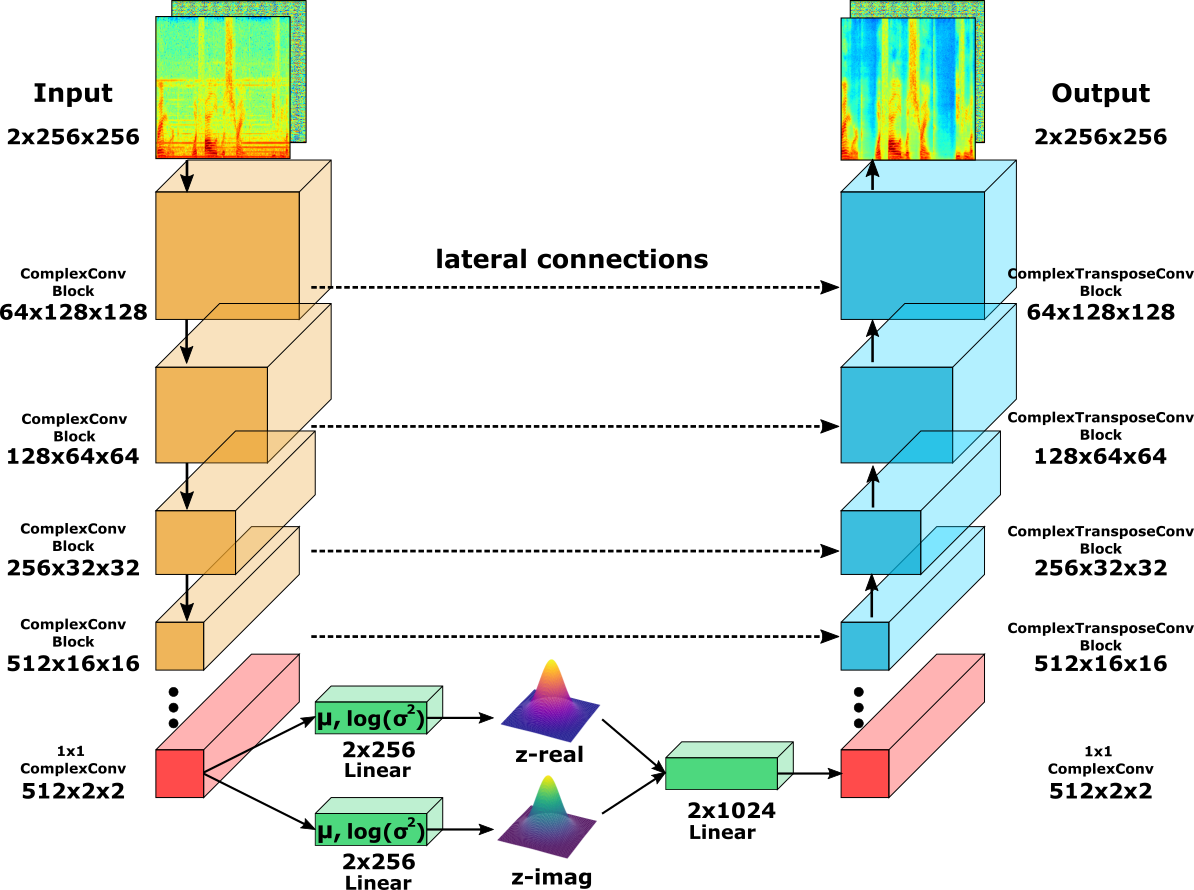}
    \caption{Schematic overview of the complex variational U-Net. The latent space (bottom) adapts a Gaussian model each for the real and imaginary parts. Vertical dots indicate two convolutional blocks (size $512\times 8 \times 8$ and $512 \times 4 \times 4$) that reduce dimensionality towards the latent space layer.}
    \label{fig:cvunet}
    \vskip-3mm
\end{figure}

\begin{figure}[t]
    \centering
    \includegraphics[scale=.75]{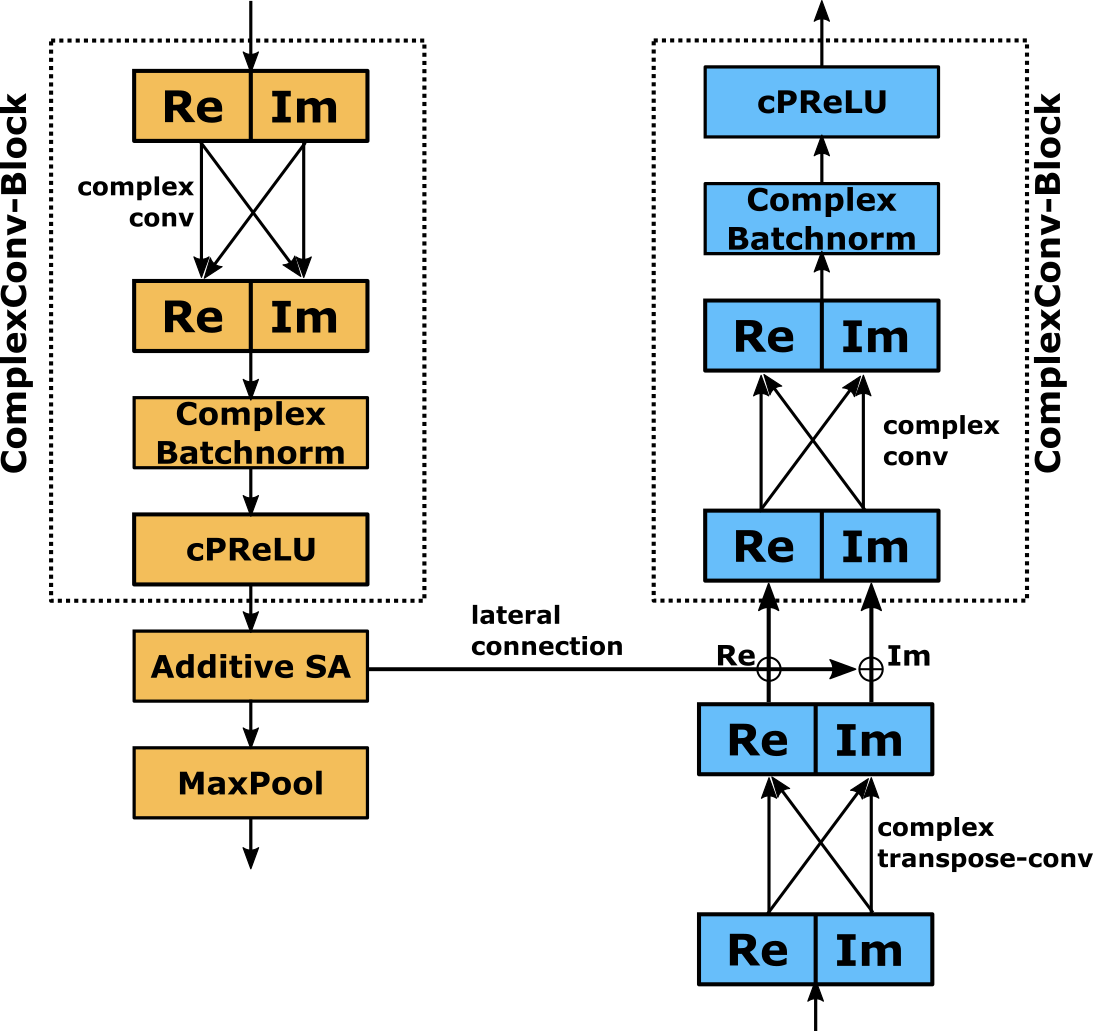}
    \caption{Detailed view of a pair of encoder (left) and decoder (right) blocks at one U-Net level. Weights in the self-attention block are computed individually for real and imaginary parts. The ``$\bigoplus$'' symbol in the decoder denotes the concatenation of respective feature matrices.}
    \label{fig:block}
    \vskip-3mm
\end{figure}

An overview of the proposed system is shown in Fig.~\ref{fig:cvunet}. Encoder and decoder paths each consist of seven convolution blocks, while the bottleneck starts and ends with a 1-by-1 convolution coupled with a complex parametric ReLU activation (cPReLU). In each encoder and decoder block, as depicted in Fig.~\ref{fig:block}, a convolution block consists of a dilated complex convolution followed by complex batch normalization and cPReLU activation. One complex convolution is composed of two real-valued convolution operations, i.e., one determines the real, the other the imaginary part of the next layer. Both convolutions use the combined real and imaginary parts of the previous layer as input, as described by Hu et al. \cite{hu20g_interspeech}. When included, self-attention (SA) is applied after each encoder block, prior to activation being relayed to the decoder via lateral connections. Convolutions in the decoder path are transpose convolutions that perform learned upsampling of signal representations towards the spectrogram resolution at the output.

\subsection{Probabilistic model and loss function}
\label{ssec:probmodel}
The standard probabilistic model for the latent space in variational autoencoders (VAEs) is a Gauss distribution \cite{IntroVAE}, which we adopt for the complex variational U-Net by modeling complex feature values through two Gauss distributions in the latent space layer.
The standard loss function for VAEs combines a reconstruction term, often chosen as a mean-squared error (MSE) function, with a Kullback-Leibler (KL) divergence that regularizes the latent space model.
Our loss function $L_{\theta, \phi}$ follows the same approach and combines three reconstruction loss MSE terms, for magnitude, real and imaginary part reconstruction, respectively, as well as the average KL divergence for both Gaussian models in our latent space, and the scale-invariant signal-to-distortion-ratio (SI-SDR) \cite{Roux2019}: 
\begin{equation}
\begin{split}
L_{\theta, \phi} = & MSE_{\text{mag}} + MSE_{\text{real}} + MSE_{\text{imag}}\\
&  + \beta \, D_{\text{KL}} - \text{SI-SDR}.
\end{split}
\end{equation}
Encoder and decoder parameters are denoted as $\phi$ and $\theta$, respectively.
Each MSE reconstruction term for the respective quantity is computed according to
\begin{equation}
    MSE = \frac{1}{N}\sum_i^N{(y_{i} - \hat{y_{i}})^2},
\end{equation}
where $\hat{y}$ denotes the reconstruction and $y$ the true value. Since the complex U-Net's output are complex-valued spectral representations, direct time-domain waveform reconstruction and inclusion of the SI-SDR,
\begin{equation}
    \text{SI-SDR} = 10 \log_{10} \left( \frac{\| \frac{\hat{\y}^T \y}{\| \y \|^2} \y \|^2 }{\| \frac{\hat{\y}^T \y}{\| \y \|^2} \y - \hat{\y}\|^2} \right),
\end{equation}
into the loss function is straightforward.
Tests indicated that uniform weighting of the MSE and SI-SDR terms, and a weighting with factor $\beta=10$ of the KL adequately balance the different terms in the $L_{\theta, \phi}$ loss function.

\section{EXPERIMENTS}
\label{sec:experiments}

\label{sec:results}
\begin{table*}[t]
    \centering
    \caption{Model performance on MS-DNS dataset in anechoic and reverberant test conditions.}
    \begin{tabular}{rcccccccccccc}
        \toprule
        \multicolumn{1}{c}{ }&\multicolumn{3}{c}{Anechoic} &\multicolumn{3}{c}{Reverberant}\\
        \midrule
        Model & PESQ & STOI & SI-SDR [dB] & PESQ & STOI & SI-SDR [dB]\\
        \midrule
        SA-CVU-Net (\emph{Ma/Ph}) & 2.90 & 0.94 & \textbf{15.95} & 2.52 & 0.84 & 8.86 \\
        SA-CVU-Net (\emph{Re/Im}) & 2.61 & 0.94 & 14.23 & 2.52 & 0.86 & 9.92 \\
        CVU-Net (\emph{Ma/Ph}) & 2.83 & 0.94 & 15.49  & 2.67 & 0.85 & 10.09 \\
        CVU-Net (\emph{Re/Im}) & 2.55 & 0.93 & 13.87 & 2.67 & \textbf{0.88 }& \textbf{11.60} \\
        CU-Net & 2.67 & 0.93 & 14.52 &\textbf{ 2.71 }& 0.87 & 11.57 \\
        DVU-Net & 3.03 & 0.94 & 12.53  & 2.39 & 0.79 & 10.77 \\
        DU-Net & \textbf{3.08} & 0.94 & 12.62  & 2.44 & 0.80 & 10.98 \\
        WaveUNet & 2.75 & 0.94 & 13.83  & 2.28 & 0.78 & 6.67 \\
        \bottomrule
    \end{tabular}
    \label{tab:dns}
    \vskip-3mm
\end{table*}

\subsection{Datasets}
\label{ssec:data}
Models are trained and evaluated on two different speech enhancement corpora. All data are resampled to 16 kHz where applicable. The MS-DNS 2020 Challenge dataset \cite{reddy2020interspeech} is used to generate 100 hours of anechoic audio data with SNR distributed uniformly in the range of 0 to 20~dB for training, as well as one hour of validation data for monitoring during training. Two evaluation sets, anechoic and reverberant, are included with an SNR of 0 to 20~dB containing 20 unknown speakers and noise types, excluding speech-like noise. Note that we train exclusively on anechoic data, while evaluation uses anechoic as well as reverberant data and, thus, intentionally includes a train-test mismatch. The Voicebank+Demand corpus \cite{valentini2017} includes speech-like and babble noise. The evaluation set has an SNR ranging from 2.5~dB to 17.5~dB compared to the training set of 0 to 15~dB relative to target speech. We train on the 19 hour long 56 speaker subset, utilizing one hour of the 28 speaker subset for validation. The 34 minute long evaluation set contains two unknown speakers and five (different) noise types.

\subsection{Network configuration}
\label{ssec:config}
The input to our model consists of short-term Fourier transform (STFT) spectrograms of length 1.6~s with 256~time frames and 256~feature (frequency) bins. Two encoding schemes for the complex-valued STFT values are considered in our experiments: For real-imaginary part (``Re/Im'') encoding, the STFTs' real part forms the first input channel and their imaginary part the second channel. 
For magnitude-phase (``Ma/Ph'') encoding, the log-scaled magnitude spectrum and phase in radians form the first and second input channel, respectively.
In either case, input and output size is (2 $\times$ 256 $\times$ 256) representing (channel $\times$ time $\times$ feature). Encoder-decoder blocks on each U-Net level contain the same size of features, starting at the input/output (2 $\times$ 256 $\times$ 256) and increasing in channels (64, 128, 256, 512, 512, 512, 512) while decreasing in time and feature dimension by half in every subsequent layer. The input of the bottleneck (512 $\times$ 2 $\times$ 2) is downsampled again as variance and mean of the two diagonal Gaussians have 256 parameters each, resulting in two latent space models after reparameterization. Both are projected by linear layers (2 $\times$ 1024), then combined to recreate the input size (512 $\times$ 2 $\times$ 2). The dilation rate scales inversely with the channel size (16, 8, 4, 2, 1, 1, 1). For reference, WaveUNet and PHASEN models are employed with input and output features as described in the respective publications \cite{Phasen,Stoller2018}.

\subsection{Evaluation setup}
\label{ssec:results}
Performance evaluation of speech enhancement models is carried out with three state-of-the-art metrics, the ITU-T P.862 speech quality standard (PESQ) \cite{PESQ}, the short-time objective intelligibility measure (STOI) \cite{STOI}, and, representing a non-perceptual measure, the scale-invariant signal-to-distortion-ratio (SI-SDR) \cite{Roux2019}.
We compare several versions of the complex U-Network architecture, which differ by degree of ablation as well as by feature encoding.
Specifically, we compare the proposed model using real and imaginary part encoding of audio signals (``\emph{Re/Im}'') with an identical model working on log-scaled magnitude and phase information (``\emph{Ma/Ph}''). Model ablations, to study relevance of its components, are introduced by successively removing from the full (SA-CVU-Net) model its self-attention component (resulting in CVU-Net) and by replacing the variational latent layer with a deterministic bottleneck layer (resulting in CU-Net with magnitude/phase encoding). Our previously proposed denoising variational U-Net (DVU-Net) and its non-variational version (DU-Net) are compared, as well. DVU-Net uses log-scaled magnitude spectra inputs without phase information. DU-Net has a deterministic latent space.

\section{RESULTS}
\begin{figure}[t]
    \centering
    \includegraphics[width=1\linewidth]{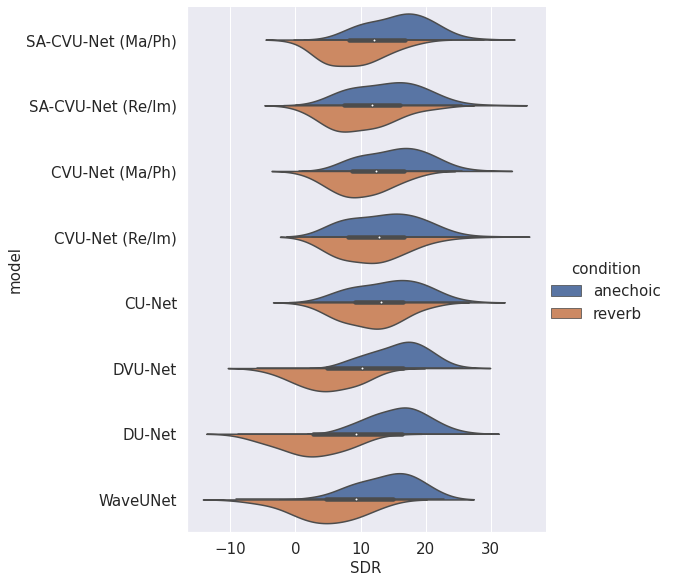}
    \caption{Distribution of SI-SDR scores (dB) for all models on the DNS-Challenge 2020 dataset. The evaluation compares the performance on unknown noise types and speakers in anechoic and reverberant conditions.}
    \label{fig:dns_eval}
    \vskip-3mm
\end{figure}

Model performance for both datasets is reported in Tab.~\ref{tab:dns} for the DNS Challenge dataset and Tab.~\ref{tab:voice} for Voicebank+Demand. Distribution of SI-SDR values for all evaluation samples of the DNS challenge is depicted in Fig.~\ref{fig:dns_eval}, directly comparing anechoic and reverberant conditions. In the anechoic scenario, the SA-CVU-Net \emph{(Ma/Ph)} achieves the highest SI-SDR score of 15.95 dB, followed by the CVU-Net \emph{(Ma/Ph)}. Both \emph{(Re/Im)} variants score 1.62 to 1.72 dB lower. In both cases, self-attention increases SI-SDR by 0.36 to 0.46 dB, with similar improvements in PESQ and STOI.
In reverberant conditions, the \emph{(Re/Im)} models perform significantly better than their counterparts, by up to 1.51 dB for the CVU-Net with a final score of 11.60 dB, while only losing about 2 dB compared to anechoic conditions. Here, self-attention is detrimental to model performance, reducing it by 1.68 dB SI-SDR. The CU-Net shows similar numbers to CVU-Net \emph{(Re/Im)}, about 3 dB less than in anechoic scenarios, followed by DVU-Net, DU-Net, and WaveUNet.
For the Voicebank+Demand corpus, CVU-Net \emph{(Mag/Ph)} scores the best SI-SDR of 20.21 dB, followed by SA-CVU-Net with 19.88 dB. CU-Net shows the third highest SI-SDR at 19.70 dB, followed by the corresponding \emph{(Re/Im)} models. In comparison, using real and imaginary part input encoding yields a 0.66 and 0.39 dB lower SI-SDR, for CVU-Net and SA-CVU-Net, respectively. Self-attention diminishes achieved scores for both input variations. WaveUNet achieves 17.83 dB SI-SDR, followed by DVU-Net at 15.62 dB, PHASEN with 13.49 dB, and DU-Net with 11.44 dB SI-SDR. Although low in SI-SDR, PHASEN reports the highest PESQ score of 3.73, the second highest being 3.37 by CVU-Net \emph{(Ma/Ph)}. 

\section{DISCUSSION \& CONCLUSION}
\label{sec:discussion}
Single-channel speech enhancement benefits heavily from incorporating phase information into the signal processing chain in one way or another, which has been shown in \cite{Phasen, hu20g_interspeech}, and also holds in our experiments. E.g., for the Voicebank+Demand corpus, DU-Net and DVU-Net achieve the lowest scores of all models. The results also show that the probabilistic bottleneck model improves performance, comparing DU-Net with DVU-Net and CU-Net with CVU-Net. Similarly, on the anechoic DNS challenge evaluation set, the \emph{(Ma/Ph)}-based CVU-Nets yield higher SI-SDR scores than CU-Net.
Processing the real and imaginary parts is significantly better than using the log-magnitude and phase spectra in reverberant conditions, but worse in anechoic. Considering that training is only done in anechoic conditions, we hypothesize that using real and imaginary parts results in less overfitting on the room acoustics. 
The physical sound generation process can be viewed as inducing a semi-independence between magnitude and phase, e.g., permitting signal level amplification without affecting phase and permitting time-shifts that affect only phase but not magnitude.
Thus, network training with a magnitude-phase encoding is faster and yields better performance under anechoic conditions.
Reverberation, however, partially removes this semi-independence of magnitude and phase through the superposition of phase-shifted signal components, largely caused by room reflections. It therefore poses a generalization problem to a trained network with magnitude-phase representation.
Real and imaginary part network training, in contrast, is slower, presumably because the network learns to implicitly model the non-linear dependence between real and imaginary parts. Generalization to reverberant conditions is more robust with this learned non-linear dependence model.
The decreased efficiency in reverberant conditions for self-attention models might plausibly be due to overfitting, too, as self-attention nearly doubles network size and increases the focus on salient features which are present during training, i.e., in anechoic conditions.
\begin{table}[t]
    \centering
    \caption{Evaluation results on Voicebank-Demand corpus.}
    \begin{tabular}{rccc}
        \toprule
        Model & PESQ & STOI & SI-SDR [dB]  \\
        \midrule
        SA-CVU-Net (\emph{Ma/Ph}) & 3.36 & 0.95 & 19.88  \\
        SA-CVU-Net (\emph{Re/Im}) & 3.31 & 0.95 & 19.49 \\
        CVU-Net (\emph{Ma/Ph}) & 3.37 & 0.95 & \textbf{20.21}\\
        CVU-Net (\emph{Re/Im}) & 3.23 & 0.95 & 19.55 \\
        CU-Net & 3.34 & 0.95 & 19.70\\
        DVU-Net & 3.32 & 0.93 & 15.62 \\
        DU-Net & 2.86 & 0.87 & 11.44\\
        WaveUNet & 3.03 & 0.72 & 17.83\\
        PHASEN & \textbf{3.73} & 0.95 & 13.49 \\
        \bottomrule
    \end{tabular}
    \label{tab:voice}
    \vskip-3mm
\end{table}
\newline
In conclusion, processing representations of complex spectra increases model performance considerably. Adding probabilistic latent space models can improve performance, but depends on the architecture and selected test conditions. Specifically, the probabilistic latent space model successfully increases performance tested against unknown noise types and speakers, without significant impact for severe differences in room acoustics like reverberation. Further, evidence is obtained that direct processing of real and imaginary parts of a signal can increase the adaptability in reverberant conditions compared to selecting magnitude-phase features. Consequently, we should be able to combine both features to achieve high performance in anechoic conditions with strong adaptability for differing room conditions like reverberation.

\bibliographystyle{IEEEtran}
\bibliography{refs}

\end{document}